\documentclass[showpacs,showkeys,preprintnumbers,amsmath,amssymb,latexsym,floatfix,superscriptaddress,prl,twocolumn]{revtex4-2}
\usepackage{lineno}
\usepackage{graphicx}
\usepackage{dcolumn}
\usepackage{bm}
\usepackage{nicefrac}
\usepackage{subfigure}
\usepackage{xcolor}
\usepackage{textcomp}

\usepackage[english]{babel}

\usepackage{amsmath}
\usepackage{graphicx}
\usepackage[colorlinks=true, allcolors=blue]{hyperref}

\begin{document}
\title{The Remodeling of Fiber Distributions in Biological Tissues:\\ Rotation without Rotation}
\author{C. Cherubini}
\affiliation{Department of Science and Bio-Technology, Università Campus Bio-Medico di Roma, Via Álvaro del Portillo 21, 00128 Rome, Italy}
\affiliation{ICRANet, Piazza della Repubblica 10, 65122 Pescara, Italy}

\author{M. Vasta}
\affiliation{Department of Engineering, Università Campus Bio-Medico di Roma, Via Álvaro del Portillo 21, 00128 Rome, Italy}
\affiliation{Department of Engineering and Geology, Università degli Studi “G. d’Annunzio” Chieti–Pescara, Viale Pindaro 42, 65127 Pescara, Italy}

\author{F. Recrosi}
\affiliation{Department of Engineering and Geology, Università degli Studi “G. d’Annunzio” Chieti–Pescara, Viale Pindaro 42, 65127 Pescara, Italy}

\author{A. Gizzi}
\email{a.gizzi@unicampus.it}
\affiliation{Department of Engineering, Università Campus Bio-Medico di Roma, Via Álvaro del Portillo 21, 00128 Rome, Italy}

\begin{abstract}
Collagen remodeling in living tissues exhibits anisotropic orientation patterns commonly described by Von Mises distributions, yet the physical origin of such nonequilibrium organization remains unresolved. In the present work, we demonstrate analytically that the combined action of Malthusian growth dynamics and the introduction of linear relations governing mechanical remodeling naturally gives rise to generalized bimodal Von Mises distributions as emergent states of living matter. The theory reveals a {\it rotation without rotation} mechanism, in which fibers progressively reorient in the absence of angular mechanical coupling via selective deposition and removal along preferred directions. The resulting analytical solutions quantitatively reproduce experimentally observed distributions and establish a direct mechanobiological origin for directional statistics in biological tissues. By interpreting the evolving normalized fiber density as a probability distribution function, we formulate a dynamical Shannon entropy framework that captures the temporal emergence of microstructural organization. The theory further yields closed-form expressions for the drift of the associated Fokker--Planck equation, enabling the corresponding stochastic differential equation to be derived, thus revealing that tissue remodeling is the collective outcome of noisy single-fiber dynamics. These results establish a minimal theoretical framework that connects biomechanics, stochastic processes, and nonequilibrium statistical organization in living matter.
\end{abstract}

\maketitle
\paragraph{Introduction.}
Collagen ($\sim10,\rm{nm}$) is the most abundant protein in the human body, playing a fundamental role in cellular function and the maintenance of tissue architecture~\cite{LOBUGLIO}. Polarized light microscopy experiments reveal a variety of directional patterns associated with the orientation dispersion of Collagen fibers ($\sim10 \,\rm{\mu m}$) in load-bearing soft biological tissues, which are commonly characterized as families of Von Mises (VM) circular distributions~\cite{ETTEMA,UEDA,TURCA, Holz,Holz2}. However, despite extensive experimental investigation, theoretical questions still remain, particularly in the context of disorder-to-order dynamics and entropy evolution. This Letter, therefore, revisits and extends a recently proposed statistical fiber remodeling framework~\cite{GIZZI}, based on the homogenized constrained mixture theory \cite{CYRON}. We analytically identify the minimal ingredients required for the generation of VM fiber patterns, namely, through the combined action of first-order Malthusian dynamics and linear mechanical remodeling kinetics. The presented theory predicts that the rotational evolution of fiber density (a polarized growth) is driven by mechanical stretching, even in the absence of angular spatial gradients ({\it rotation without rotation}). Interpreting these results in terms of a probability distribution function (PDF), governed by a Fokker–Planck equation, permits the formulation of a dynamical Shannon entropy framework that is asymptotically associated with increasingly organized tissue microstructures. The arrived-at PDF formulation further allows for the introduction of a companion stochastic differential equation, reframing remodeling as a noisy single-fiber process from which the continuum theory can be recovered via Monte Carlo simulations. 
\begin{figure}{}
    \subfigure[]{\includegraphics[width=0.22\textwidth, clip]{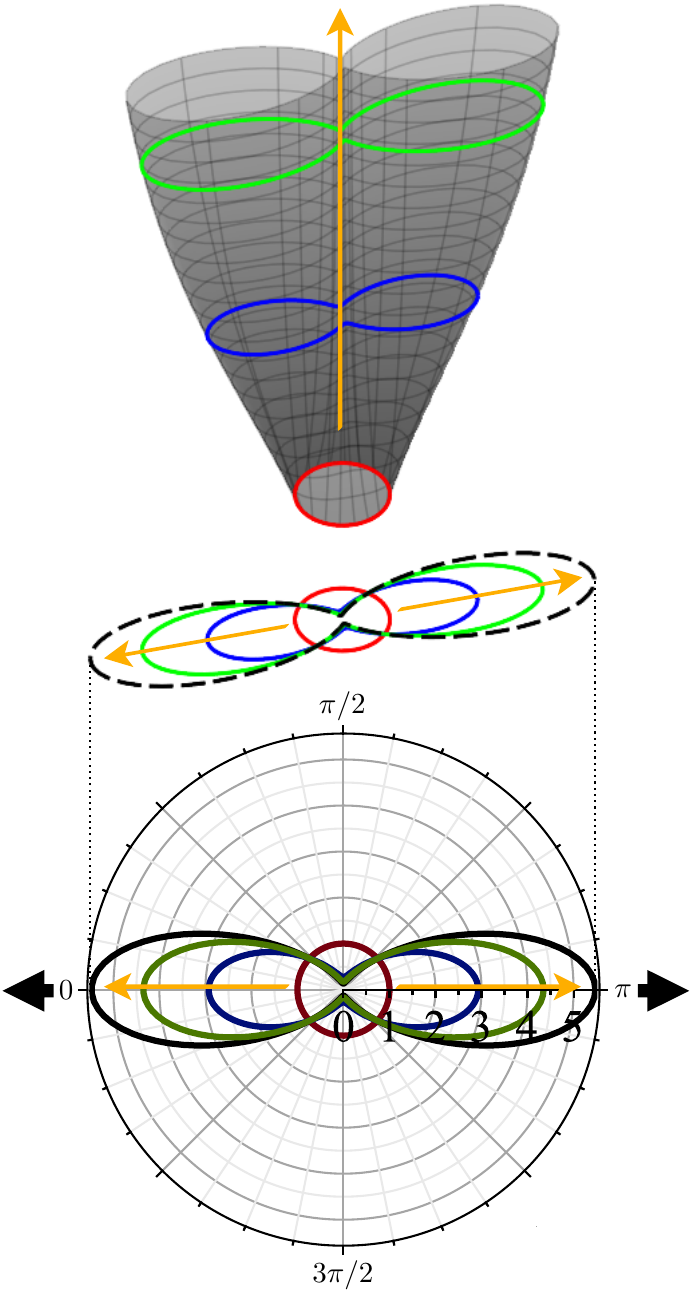}
    \label{fig:polarU}}
    \subfigure[]{\includegraphics[width=0.22\textwidth, clip]{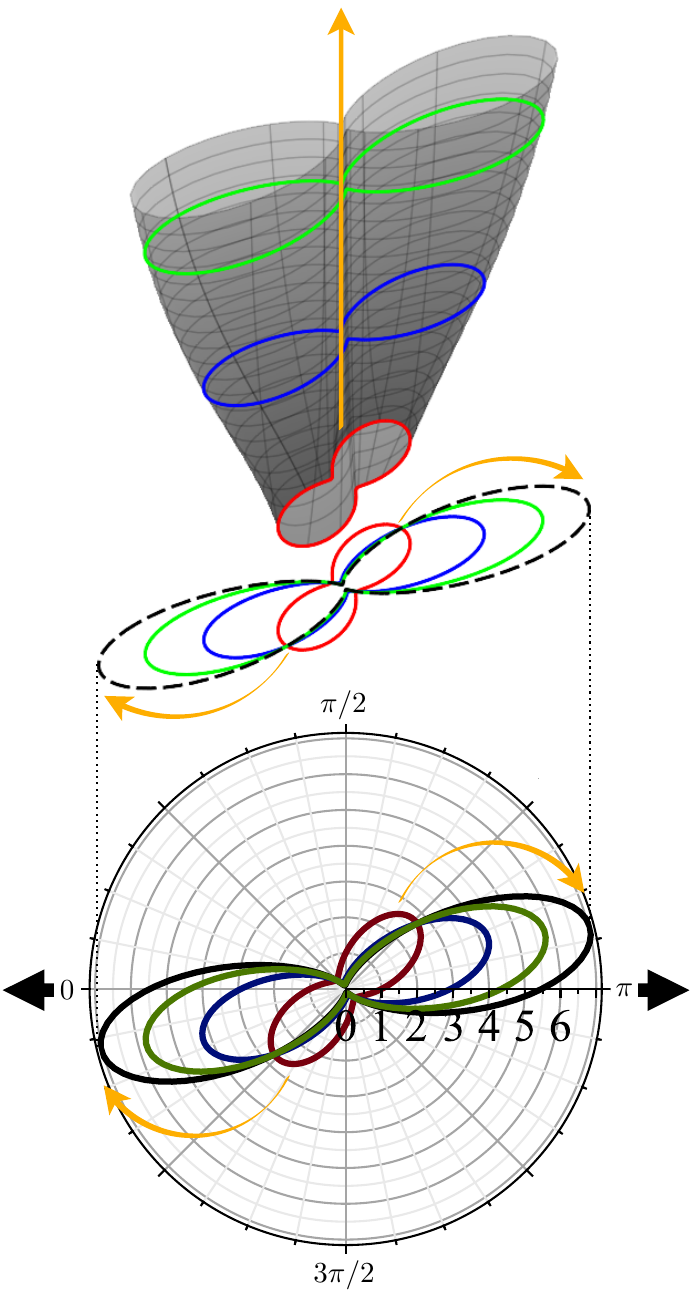}
    \label{fig:polarD}} 
    \caption{Fiber density polar plot evolution subject to external loading at $t=0,5,10,+\infty$. (a) Initial circular distribution elongates along the loading axis; (b) initial BVM distribution rotates and elongates around the loading axis. Asymptotic configurations (black) represent different transversely isotropic material models.
    Upper panels: companion spacetime diagram for the isochronous contours (time as yellow axis), a visualization tool typical of String Theory~\cite{Zwie}, highlighting (a) the quasi-bifurcation in time {\it world-sheet}, and (b) twisting during evolution.
    Model parameters $\lambda = 1.2, \tau_{I_4} = 5, \tau_\rho = 1, I_{4_{e,h}} = 1.1, \rho_0 = 1$ and $b_0=1,\theta_0=\pi/4$.
    \label{fig:polar}}  
  \end{figure}

\paragraph{Modeling.} 
We refer to Gizzi et al.~\cite{GIZZI} for the homogenized constrained-mixture mechanical framework. In the presented work, we specialize the formulation to a planar problem representative of a homogeneous tissue embedded with distributed collagen reinforcement. The time-varying collagen density along a generic angular direction is denoted by $\rho=\rho(t,\theta)$, with $\theta\in[0,2\pi]$, and describes a bundle of fibers capable of supporting both tension and compression. Each value of $\theta$ identifies a specific fiber-bundle orientation and, consequently, a distinct constituent family within the mixture. Accordingly, the total fiber mass density $\rho_T(t)$ and the total fiber mass $M$ contained in a volume $V$ are given by:
\begin{equation}
\rho_T(t)=\int_{0}^{2\pi}\rho(t,\theta')d \theta' \,,\quad
M=\int_V\rho_T dV \,,
\end{equation}
with $\rho_T=dM/dV$. Therefore, $\rho=d \rho_T/d\theta$ represents the mass density of a given number of fibers (each of constant mass $m_0$) contained within an infinitesimal volume $dV$ and aligned along a fixed direction $\theta$. 
In a minimal remodeling framework, the dynamics are assumed to evolve independently along each selected angle $\theta$, thus implying that fibers are non-interacting. Angular derivatives are therefore excluded, and the angle $\theta$ appears only as a parameter in the analysis. As a starting point for deriving a constitutive evolution equation for the fiber density, we assume that $\rho(t,\theta)$ follows non-autonomous first-order Malthusian growth kinetics~\cite{Murray1}, such that:
\begin{equation}\label{eq:RhoDin}
    \frac{d\rho}{dt}
    =
    \rho {\mathbf K}:\left( {\mathbf S}-{\mathbf S}_h\right)\,,
\end{equation}
where the evolution of fiber density is driven by the difference between the anisotropic second Piola-Kirchhoff stress tensor ${\mathbf S}={\mathbf S}(I_4)$ and its preferential homeostatic value ${\mathbf S}_h$~\cite{Miller}, motivated by the homeostatic stress hypothesis~\cite{Fung}. Here, $I_4={\mathbf C}:{\mathbf A}$ represents the fourth pseudo-invariant of strain, where $\mathbf C$ is the right Cauchy-Green deformation tensor and ${\mathbf A}\equiv \vec n \otimes \vec n $ is the fiber's structural tensor, with $\vec n=(\cos\theta,\sin\theta,0)$ denoting the fiber's unit direction vector. Lastly, $\mathbf K$ is a gain constitutive tensor that is defined subsequently.

Biomechanical evidence indicates that the anisotropic contribution of the strain energy density associated with fiber reinforcement follows an exponential law $\psi(I_{4})=c_1 [e^{c_2(I_{4}-1)^2}-1]$, where $c_1$ and $c_2$ are real constants.
Applying the thermodynamics-based Coleman-Noll procedure, we obtain: 
\begin{equation}
    {\mathbf S} =
    4 c_{1}c_{2}(I_4-1)e^{c_2(I_4-1)^2} {\mathbf A}
    \equiv F(I_4) {\mathbf A}\,.
\end{equation}
For a fixed $\theta$-direction, Eq.~\eqref{eq:RhoDin} then becomes 
\begin{equation}
\label{eq:RHO_F}
    \frac{d\rho}{dt}
    =
    G\left[F(I_4)-F(I_{4_h})\right]\rho\,,
\end{equation}
with $G=G(t,\theta)={\mathbf K}:{\mathbf A}$. Consistent with a minimal formulation, we consider the simplest scenario with ${\mathbf K}=k_{0} {\mathbf I}$ and $k_0=const$, such that $G\equiv k_0$. This choice is physically sound, since a uniform planar deformation $\lambda$ (identical in both the vertical and horizontal directions) results in a homothetic transformation that preserves isotropy. Thus, the rate of Malthusian dynamics is mechanically linked to the trace of the second Piola-Kirchhoff tensor, an invariant quantity of the theory. 

Performing a first order Taylor series expansion of Eq.~\eqref{eq:RHO_F} around $I_4 = I_{4_h}$ yields ${d\rho}/{dt}=T^{-1}(I_4-I_{4_h})\rho$, where $T$ is the characteristic time scale of the remodeling process. The generalised framework also accounts for finite kinematics by introducing the biologically motivated multiplicative decomposition of the deformation gradient $\mathbf F=\mathbf F_e \mathbf F_{g}$ into an elastic part, $\mathbf F_e$, and an inelastic growth part, $\mathbf F_{g}$. Accordingly, the fourth pseudo-invariant follows a similar decomposition, $I_4=I_{4_e}I_{4_{g}}$ while the homeostatic configuration also satisfies $I_{4_h}=I_{4_{e,h}}I_{4_{g,h}}$. 
Guided by the physical arguments and theory presented in~\cite{GIZZI}, we postulate that the fiber density dynamics, considering only small perturbations away from homeostasis, are governed solely by the elastic part of the deformation. As such, instead of the Taylor series expansion of Eq.~\eqref{eq:RhoDin}, introduced previously, we arrive at:
\begin{equation}
\label{eq:ODE_RHO}
    \frac{d \rho}{dt}
    =
    \left(\frac{ I_{4_e}(t,\theta)-I_{4_{e,h}}}{\tau_{\rho}}\right)\rho \,\,,\,\,
    \rho(0,\theta)
    =
    \rho_{0}(\theta) \,,
\end{equation}
where $\tau_\rho$ is the associated time constant. Additionally, we consider the elastic fourth invariant $I_{4_e}$ to develop subject to a first-order time-dependent process defined by the relation~\cite{GIZZI}:
\begin{equation}
\label{eq:ODE_I4}
    \frac{d I_{4_e}}{dt}=\left(\frac{I_{4_{e,h}}- I_{4_e}(t,\theta)}{\tau_{I_4}}\right)\,,\,\,
    I_{4_e}(0)
    =
   { I_{4_{e0}}(\theta)\,,}
\end{equation}
where again, $\tau_{I_{4}}$ is a characteristic time constant.

\paragraph{Uniaxial loading.}
We now consider a fixed uniaxial isochoric elongation along the $x$-axis defined by the macroscopic dimensionless stretch ratio $\lambda$, such that ${\bf C}={\rm diag}(\lambda^2,1/\lambda^2,1)$. As the fourth invariant $I_4$ is constant in time, we find that $I_{4_{g}}(t)=I_4/I_{4_e}(t)$. Additionally, the elastic fourth pseudo-invariant is initialized to $I_{4_e}(t=0,\theta)
\equiv
I_{4_{e,0}}(\theta)
$, 
and, as growth is not active at $t=0$, the growth-related fourth pseudo-invariant is set to $I_{4_g}(0)=1$. Accordingly, we observe that:
\begin{equation}
   I_{4_{e,0}}(\theta)
    =
    \lambda^2\cos^2\theta+\frac{1}{\lambda^2}\sin^2\theta
    \equiv 
    S_1\cos2\theta+S_2\,,
\end{equation}
with 
$S_1={\nicefrac12\left(\lambda^2-{\lambda^{-2}}\right)}$, and
$S_2=\nicefrac12\left(\lambda^2+{\lambda^{-2}}\right)$,
where $S_{1}$ and $S_{2}$ are dimensionless geometrical quantities. The appearance of the term $\cos 2\theta$ demonstrates an angular bi-periodicity, conforming with the tensorial representation of a fiber, i.e., two collinear vectors. Consistent with such an outcome we consider the fibers directed along $\theta$ and $\theta+\pi$ to be two distinct (although collinear) entities.

The solution to Eq.~\eqref{eq:ODE_RHO} and Eq.~\eqref{eq:ODE_I4} for the considered uniaxial problem then becomes:
\begin{eqnarray}
    I_{4_e}(t,\theta)&=&I_{4_{e,h}}+e^{-\frac{t}{\tau_{I_4}}}\left( S_1\cos 2\theta+S_2-I_{4_{e,h}}\right)\,,\nonumber
    \\
    \rho(t,\theta)&=&\tilde \rho_{o}(\theta)A(t)e^{b(t)\cos 2\theta} \,,
\end{eqnarray}
with the various terms defined as
\begin{eqnarray}
    A(t)
    &=&
    \exp\left[{\frac{\tau_{I_4}}{\tau_{I_\rho}}(S_2-I_{4_{e,h}}) f(t)}\right] \,,\nonumber
    \\
    b(t)&=&\Omega f(t) \,,\nonumber
    \\
    \Omega &=&\nicefrac{\tau_{I_4}}{\tau_{I_\rho}} S_1\equiv {\cal R} S_1 \,,\nonumber
    \\
    f(t)&=&1-\exp\left[{-\nicefrac{t}{\tau_{I_4}}}\right] \,.
\end{eqnarray}
It is worth noting the emergence of the key dimensionless group $\Omega$, which links the ratio $\mathcal{R}$ of the two characteristic time scales to the macroscopic dimensionless stretch via $S_1 = S_1(\lambda)$. Throughout the subsequent analysis we set the cellular remodeling time scale to unity, $\tau_{\rho} = 1$, such that time is measured in units of this reference scale.

\paragraph{Rotation without rotation.}
In the previous section we demonstrated that a time-dependent generalized bimodal Von Mises (BVM) function~\cite{Mardia}, $A(t)e^{b(t)\cos 2\theta}$ (where $b(t)$ is commonly referred to as the {\it concentration parameter}), depending solely on the parameters $\Omega, \lambda,\tau_{I_4}$, is the fundamental driving mechanism governing the evolution of any initial density configuration $\tilde \rho_{0}(\theta)$. 
For instance, if an initial uniform fiber density is considered, $\tilde \rho_{0}(\theta)=\rho_{0}$, the circular distribution naturally evolves into an $x$-axis symmetric BVM function, becoming progressively more concentrated with increasing time, and characteristically more transversely isotropic in nature as $t\to+\infty$, see Fig.~\ref{fig:polarU}. However, if the initial configuration is instead circularly asymmetric, i.e., $\tilde \rho_{0}(\theta)=\rho_0 \exp[{b_0\cos2(\theta-\theta_0)}]$, where $\theta_0$ is the initial angle of orientation, the distribution rotates in time towards the loading $x$-axis tending to a BVM asymptotically aligned around a new direction, see Fig.~\ref{fig:polarD}. We term this phenomenological outcome “rotation without rotation,” as the formulation lacks spatial derivatives; nevertheless, an effective $\theta$-dynamics emerges from the biomechanical metabolism (elimination and deposition) of collagen along each angular direction. Crucially, the arrived at analytical laws of evolution imply that an externally imposed stretch can mechanically steer the direction of the remodeling process, in line with mechanobiological expectations~\cite {Humphrey}. 
\begin{figure*}
\includegraphics[width=1\linewidth]{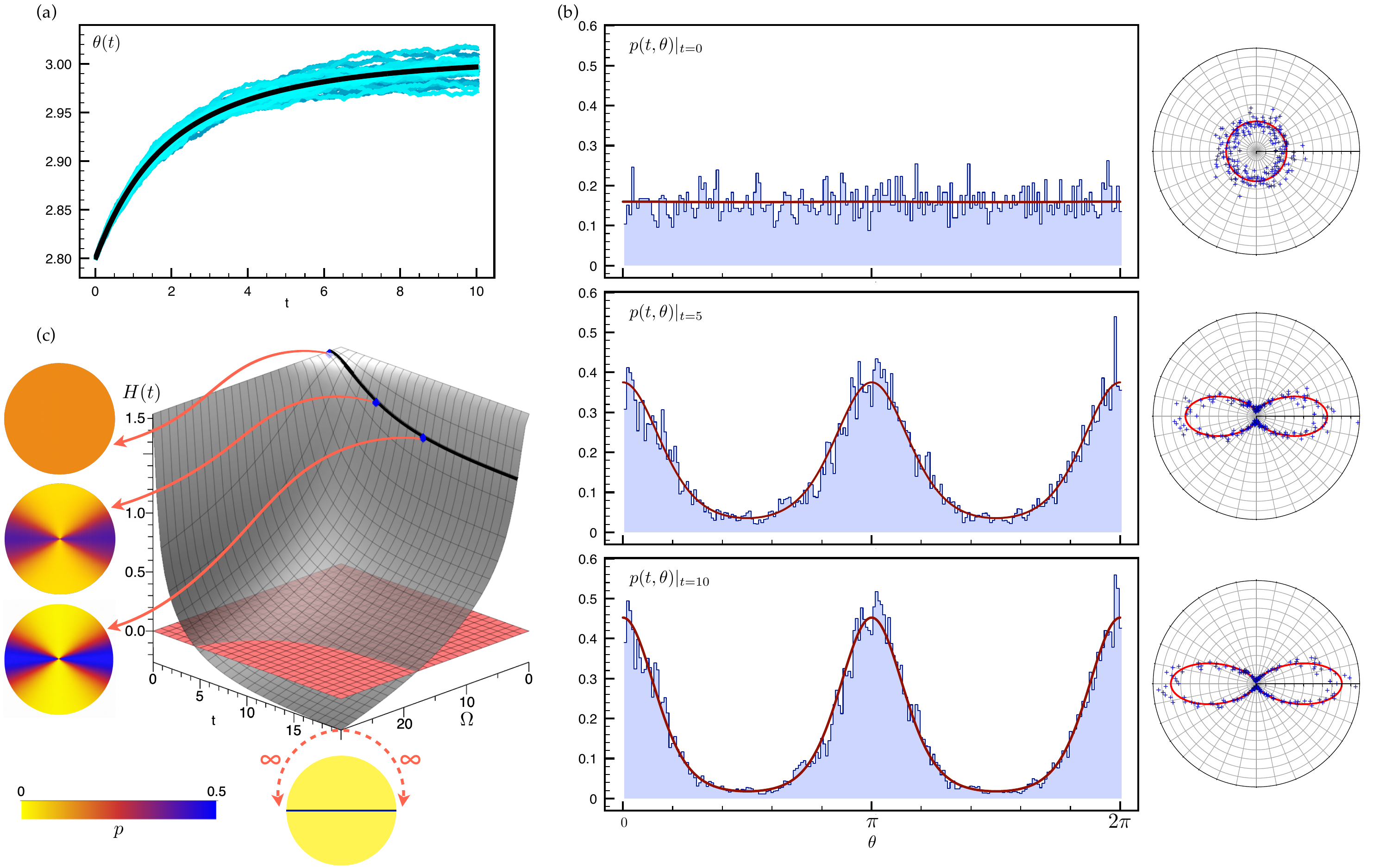}
    \caption{
    (a) 
    Time course for an initial angle, $\theta(t=0)=2.8$, comparing the deterministic outcome (solid black) with fifty stochastic trajectories (blue), showing a fluctuation around a preferred asymptotic direction $\theta\simeq 3.03$. 
    All simulations relate to the model parameters listed in Fig.\ref{fig:polar} and assume $R=1$ and $D=10^{-5}$ 
    (Maple V\textregistered {\it Finance} with $\Delta t=10^{-2}$).
    (b)     Binning frequency histograms (frequency ${\cal F}_i(t)$, $i=1..N$, $N=200$) 
    superimposed to the corresponding analytical PDF ($p$-red) at three representative times: the SDE was integrated for $4\cdot 10^3$ uniformly distributed initial conditions as in (a). Equivalent polar plots are shown on the right.
   (c)
    3D Shannon entropy plot, $H(t)$, for a uniform initial PDF with $\tau_{I_{4}}=5$. The surface $H=0$ (red) highlights the negative transition. 
    The solid black curve again corresponds to parameters listed in Fig.~\ref{fig:polar}. The angular probability distribution density $p=p(t,\theta)$, plotted in a manner analogous to orbitals in Quantum Mechanics~\cite{Griffiths}, is shown for selected times $t=0,5,10$ with companion binning frequency entropy $\tilde{H}(t)=-\frac{2\pi}{N}\sum_{i=1}^{N}{\cal F}_i(t) \log {\cal F}_i(t)$ (blue diamonds).  
    In the limits $\Omega\to +\infty \cup\, t\to +\infty$, a bimodal Dirac-delta type density distribution is obtained such that $H\to-\infty$ (the highest ordered configuration possible). 
    } 
    \label{fig:compact} 
    \end{figure*}
    
\paragraph{Probabilistic interpretation.}
 Biomechanical experiments rely on optical statistical analyses across multiple samples exhibiting VM-type distributions. Within our framework, a high fiber mass density $\rho$ along a given direction $\theta$ corresponds to a high concentration of matter; consequently, collagen deposition driven by probabilistic mechanisms is favored along that direction, and the associated probability density $p$ is correspondingly enhanced at angle $\theta$. This observation motivates a statistical interpretation of the BVM fiber mass-density function $\rho(t,\theta)$, normalized by $\rho_T(t)$, namely $p=p(t,\theta)={\rho(t,\theta)}/{\rho_T(t)}$ as a PDF, or, in the original deterministic mechanical context, as the fraction of fibers oriented along the direction $\theta$. Such a PDF is consistent with a stochastic process governed by a circular Fokker-Planck equation (FP) without jumps~\cite{Gardiner}. In general, for a 3D-FP equation, $p$ satisfies $\partial_t  p +\nabla\cdot( {\bf v} \rho)=D\nabla^2\rho$ where $D$ is assumed to be a constant for the sake of simplicity. We must formulate the FP on a circle of constant radius $R$, which necessitates working in cylindrical or spherical coordinates. Accordingly, the drift vector $\bf v$ is expressed with respect to the corresponding orthonormal curvilinear frame, on which the divergence and the Laplacian operators are also defined. Starting from the case of a uniform initial fiber distribution, $\tilde \rho_{0}(\theta)=\rho_0$, the normalized PDF takes the form $p(t,\theta)=\exp{{(b(t)\cos 2 \theta}})/(2\pi  I_0(b(t)))$, 
where $I_0(\cdot)$ is a modified Bessel function of order zero. By imposing a constant radius $r=R$ and restricting the analysis to the equatorial plane (or equivalently considering constant radius $R$ and $z=0$ in cylindrical coordinates), we obtain a curvilinear diffusion-advection PDE \cite{Krause,Brillinger,Lengerich} which collapses into the desired circular FP:
\begin{equation} 
\label{eq:circularFPE}
    \frac{\partial p}{\partial t}
    =
    -\frac{1}{R}\frac{\partial \left( p v_\theta\right)}{\partial \theta}+\frac{D}{R^2}\frac{\partial^2 p}{\partial \theta^2} \,.
\end{equation}
Given that $p(t,\theta)$ is coincident with the aforementioned normalized BVM, we compute the unknown drift by solving a linear ordinary differential equation for $v_\theta$, subject to the initial condition $v_\theta(t,0)=0$, implying that fibers aligned with the positive real $x$-axis experience no drift. The proposed procedure addresses the well-known inverse Fokker–Planck problem~\cite{Inverse}, which is also relevant to other fields such as economics (e.g., the {\it smile} problem~\cite{SMILE1}). It is worth noting that, in the present Letter, we substantially extend the framework proposed in~\cite{GIZZI} by introducing a constant diffusion coefficient $D$, representative of intrinsic biological noise: 
\begin{eqnarray}
\nonumber
v_\theta(t,\theta)
&=&
-\frac{R}{p}\int_{0}^{\theta}\frac{\partial}{\partial t} p(t,\theta')d\theta'+
\frac{D}{R}\frac{\partial}{\partial \theta}\log p
\\
&\equiv&
-
\frac{R}{p}\frac{\partial}{\partial t}\Phi(t,\theta)-\frac{2D}{R}b(t)\sin 2\theta \,,
\end{eqnarray}
where, $\Phi(t,\theta)=\int_{0}^{\theta}\rho(t,\theta')d\theta'$ 
is the cumulative distribution function  of the BVM PDF, whose explicit form can only be obtained as an infinite series. In the limit $t\to +\infty$, the only non-vanishing term in the drift is the one driven by the diffusion coefficient, yielding $v_{\theta,\infty}(\theta)=-(2D\Omega/R)\sin 2\theta$, while the asymptotic PDF becomes $p_\infty( \theta)=\exp({\Omega\cos 2\theta})/(2\pi I_0(\Omega))$. This result is consistent with the classical work of Kent~\cite{Kent}, which states that, for a constant diffusion coefficient and a drift proportional to a sine function, the mono-periodic Von Mises distribution is an exact steady-state solution of the FP equation on a circle.  We have shown that the result holds also in the case of a bi-periodic Von Mises PDF.

\paragraph{Information entropy.}
A PDF allows one to invoke Information Theory and introduce Shannon differential entropy ($I_1(\cdot)$ is a modified Bessel function of order one): 
\begin{eqnarray}
H(t)&=&-\int_{0}^{2\pi}p \log p \,d\theta=\nonumber\\
&=&\log(2\pi I_0(b(t)))-b(t)\frac{I_1(b(t))}{I_0(b(t))}\,,
\end{eqnarray}
which corresponds to the well-known entropy of the Von Mises PDF in directional Statistics~\cite{Mardia}, here extended to the time-dependent BVM case. The above expression is obtained by integrating the Fourier series expansion of $p$ together with the orthogonality properties of trigonometric functions. In the limit $t\to 0$, one recovers $H(0)=\log 2\pi$, corresponding to the maximum entropy value for a VM distribution, namely, the circular distribution, which exhibits no preferred direction and represents the lowest possible level of organization. Moreover, $H(t)$ can be expressed solely in terms of the parameters $\Omega$ and $\tau_{I_{4}}$ (the same clearly applies to $p(t,\theta)$). Consequently, in the limit $t\to +\infty$, one obtains $H\to \log(2\pi I_0(\Omega))-\Omega{I_1(\Omega)}/{I_0(\Omega)}$, which displays a monotonically decreasing behavior with respect to the dimensionless parameter $\Omega$ and may attain both positive and even negative values. A negative Shannon entropy is the signature of increasingly peaked--``more confined''--distributions~\cite{Shannon}.  Mathematically, whenever $p>1$, the logarithm becomes positive, such that negative contributions to the entropy integral may arise and eventually dominate the entire integral itself. As $t\to+\infty$, the asymptotic distribution becomes $p_\infty(\theta)=e^{\Omega\cos 2\theta}/(2\pi I_0(\Omega))$ which, in the limit of a large a BVM concentration parameter $\Omega\to+\infty$, converges to a bimodal Dirac distribution function characterized by zero uncertainty. 
In this limit, the entropy satisfies $H\to-\infty$, indicating that a perfectly unidirectional fiber material—achievable artificially but not spontaneously realized in nature—attains the minimum entropy state and, consequently, the highest degree of organization in the Shannon sense. Within the biomechanical context, such behavior underlies the differentiation and organization of the tissue microstructure in response to externally imposed loads. The system is, in fact, thermodynamically open, continuously exchanging energy and mass in order to achieve and maintain a prescribed homeostatic state~\cite{Humphrey}.

Figure~\ref{fig:compact}(c) shows the temporal evolution of the entropy subject to varied model parameters and compares it with the results obtained from the stochastic-process analysis, discussed subsequently.

\paragraph{Stochastic process dynamics.}
According to the theory of Stochastic Processes, a PDF must be related to the finite transition probability $p(t,\theta\vert t=0,\theta')$ and to the uniform initial distribution $p(t=0,\theta)=1/(2\pi)$ as:
\begin{eqnarray}
p(t,\theta)&=&\int_0^{2\pi}p(t=0,\theta')p(t,\theta\vert t=0,\theta')d\theta'=\nonumber \\&&\int_0^{2\pi}\frac{1}{2\pi} p(t,\theta\vert t=0,\theta')d\theta' \,.
\end{eqnarray}
The associated nonlinear stochastic differential equation (SDE), formulated on a local orthonormal frame for a random process evolving on a circle with drift (obtained by selecting the equatorial plane of a random process with drift in spherical coordinates~\cite{Apaza}), reads:
\begin{equation}
\label{eq:SDE}
    \dot{\theta}=\frac{v_\theta(t,\theta)}{R}+\frac{\sqrt{2D}}{R}\xi_\theta\,,
\end{equation} 
where $\xi_\theta$ denotes a white-noise process defined on the interval $[0,2\pi]$\, (all quantities involving $\theta$ are considered to be $\rm mod \, 2\pi$). 
Figure~\ref{fig:compact}(a) presents the numerical analysis for the deterministic case, obtained from Eq.~\eqref{eq:SDE} by setting $D=0$, together with $50$ stochastic trajectories corresponding to $D\neq 0$, all initialized from the same value of $\theta$. The results highlight the emergence of a fluctuation band surrounding the deterministic trajectory.
The same procedure was subsequently repeated for several thousand distinct initial conditions, leading to the numerical integration of $2\cdot10^5$ SDE realizations. By employing a binning procedure based on $200$ equally spaced bins, we demonstrate that, at selected times, the resulting frequency histograms converge to the analytical bi-periodic Von Mises PDF $p(t,\theta)$, which solves the corresponding Fokker–Planck equation, see Fig.~\ref{fig:compact}(b). The analysis also allows for the computation of the associated Shannon entropy for the stochastic process, displaying excellent agreement, at selected times, with the entropy obtained analytically from the corresponding PDF, see Fig.~\ref{fig:compact}(c).

\paragraph{Conclusions.}
In summary, this Letter establishes the minimal theoretical requirements for a mathematical model capable of reproducing the experimentally observed anisotropic orientation patterns of collagen fibers through Von Mises-type distributions. The results are predominantly analytical, while numerical simulations further validate the theoretical predictions.
Malthusian-type dynamics capture the temporal rearrangement and reorientation of collagen fibers under a uniform uniaxial stretch. Notably, the presented theory provides a clear, biologically grounded interpretation of this reorganization in terms of preferential fiber deposition and removal along selected directions—what we term “rotation without rotation”—rather than through angular mechanical coupling between neighboring fibers.
The conceptual connection between the VM distribution and a PDF enables a probabilistic formulation of the problem and provides a coherent interpretation of the out-of-equilibrium nature of biological dynamics. Within this framework, we adopt an information-theoretical perspective on collagen fiber remodeling and quantify the emergence of microstructural organization throughout the remodeling process. By analytically solving the associated Fokker–Planck equation through an inverse procedure, we derive the biological drift term and relate it both to an underlying noisy stochastic dynamics and to the corresponding information entropy.

This methodology recasts biological remodeling as the collective outcome of multiple stochastic processes, thereby providing a concrete realization of the Lagrangian single-fiber formulations recently advocated in the literature~\cite{Grillo}. Importantly, the systematic procedure of associating the fiber density $\rho$ with a probability density function $p$, and consequently introducing a Shannon entropy $H$, naturally extends beyond the present setting. The same framework may be generalized to non-planar configurations or to nonlinear reaction–diffusion scenarios capable of exhibiting biologically relevant switching phenomena in physiological and pathological contexts.

\paragraph{Acknowledgments}

The authors thank C. Miller for useful
comments that improved the presentation of the results. Authors acknowledge the Italian National Group for Mathematical Physics, GNFM-INdAM. AG acknowledges the ERC Consolidator Grant support from the European Union’s Horizon Europe research and innovation programme under grant agreement No. 101170592 - MiGEM.

\paragraph{Data availability.}
The data are available from the authors upon request with no restriction.

\bibliographystyle{apsrev4-2}
\bibliography{sample}

\end{document}